\documentstyle[preprint,aps]{revtex}
\tightenlines

\def\be{\begin{equation}}
\def\ee{\end{equation}}

\def\bea{\begin{eqnarray}}
\def\eea{\end{eqnarray}}

\begin{document}
\draft
\title{Exact finite-size corrections for the square lattice Ising model\\
with Brascamp-Kunz boundary conditions}

\author{N. Sh. Izmailian,$^{1,2}$  K. B. Oganesyan,$^{1,2}$ and
Chin-Kun Hu$^{1,*}$}

\address{$^1$Institute of Physics, Academia Sinica,
Nankang, Taipei 11529, Taiwan}

\address{$^2$Yerevan Physics Institute, Alikhanian Br. 2, 375036
Yerevan, Armenia}

\maketitle


\begin{abstract}
Finite-size scaling, finite-size corrections, and boundary effects for
critical systems have attracted much attention in recent years. Here
we derive exact finite-size corrections for the free energy $F$
and the specific heat $C$ of the critical ferromagnetic Ising model on the
${\cal M} \times 2 {\cal N}$ square lattice with Brascamp-Kunz (BK) boundary
conditions [J. Math. Phys. {\bf 15}, 66 (1974)]
and compared such results with those under toroidal  boundary conditions.
 When the ratio $\xi/2=({\cal M}+1) / 2 {\cal N}$
is smaller than 1 the behaviors of finite-size corrections for $C$
are quite different for BK and toroidal boundary conditions;
when $\ln (\xi/2)$ is larger than 3, finite-size corrections for $C$
in two boundary conditions approach the same values.
In the limit  ${\cal N} \to \infty$ we obtain
the expansion of the free energy for infinitely long strip with
BK boundary conditions. Our results are consistent with the
conformal field theory prediction for the mixed boundary conditions
by Cardy [Nucl. Phys. {\bf B 275}, 200 (1986)]
although the definitions of boundary conditions in two cases are
different in one side of the long strip.
\end{abstract}

\vskip 0.2 cm

\newpage

\section{Introduction}
\label{introduction}

In the study of phase transitions and critical phenomena,
it is extremely important to  understand finite-size corrections
to thermodynamical quantities.  In experiments and in numerical studies
of model systems, it is essential to take into
account finite size effects in order to extract correct
infinite-volume predictions from the data. Finite-size scaling
\cite{Fisher,Barber,Privman} concerns
the critical behavior of systems in which one or more directions are
finite, even though microscopically large and is valuable in the analysis
of experimental and numerical data in many situations, for example, for
films of finite thickness. As soon as one has a finite system one
must consider the question of boundary conditions on the outer
surfaces or ``walls'' of the system. The systems under various boundary
conditions have the same per-site free energy, internal energy,
specific heat, etc, in the bulk limit, whereas the finite size corrections are
different. To understand the effects of boundary conditions on finite-size
scaling and finite-size corrections, it is valuable to study model systems,
such as percolation model \cite{sa94} and the Ising model
\cite{Onsager,Kaufman,ff69,McCoyWu,Brascamp}.
Therefore, in recent decades there have many investigations on
finite-size scaling, finite-size corrections, and boundary effects for
critical model systems
\cite{Blote,Cardy1,huetal,hu99,ih01,ih02,LuWu,Salas,Okabe,Janke,Ivasho,Salas1}.
Of particular importance in such studies are exact results where
the analysis can be carried out without numerical errors.

The Ising model has exact solutions on finite lattices with
many kinds of boundary conditions, including cylindrical
\cite{Onsager}, toroidal \cite{Kaufman,ff69,McCoyWu}, and
Mobius strip and Klein bottle \cite{LuWu}. This class also includes
the special boundary conditions introduced by Brascamp and
Kunz \cite{Brascamp}.  The calculation of the exact partition
function of the two-dimensional Ising model in the zero field wrapped
on the cylinder was
performed by Onsager in 1944 \cite{Onsager}. Exploiting the exactly
known partition function of the two dimensional Ising model on finite
$M \times N$ square
lattice with toroidal boundary conditions \cite{Kaufman}, Ferdinand
and Fisher \cite{ff69} computed the finite-size corrections to
the free energy, the internal energy, and the specific heat up to order
$N^{-1}$.
Recently, there has been much effort in understanding the behavior
of finite-size corrections of the free energy, internal energy and specific
heat. Izmailian and Hu \cite{ih02} and Salas \cite{Salas}
 extended the results of \cite{ff69} for the free energy and the internal
energy up to order $N^{-5}$ and for the specific heat up
to order $N^{-3}$.
Lu and Wu \cite{LuWu} obtained expressions for the
partition function of the Ising model on an quadratic lattice embedded
on a Mobius strip and a Klein bottle. They find finite-size corrections
for free energy to order $N^{-1}$.
Brascamp and Kunz \cite{Brascamp} calculated the partition
function of the Ising model on the $M \times 2 N$ square lattice
for special boundary conditions shown in Fig. 1. Recently
Janke and Kenna \cite{Janke} has calculated the finite-size corrections of
the specific heat for this  boundary condition up to $M^{-3}$ order.
Very recently, Ivashkevich, Izmailian and Hu \cite{Ivasho} provided a
systematic method to compute finite-size corrections to the partition
function and their derivatives of the Ising model on torus. Their approach
is based on an intimate relation between the terms of the asymptotic
expansion and the so-called Kronecker's double series \cite{Ivasho}
which are directly related to elliptic theta functions.
Expressing the final result in terms of
theta functions avoids messy sums (as in some earlier works) and greatly
simplifies the task of verifying the behavior of the different terms
in the asymptotic expansion under duality transformation
$M \leftrightarrow N$. Using this approach, Salas \cite{Salas1} computed
the finite-size corrections to the free energy, internal energy and specific
heat of the critical Ising model on a triangular and honeycomb lattices
wrapped on a torus.

Using the exact partition of Ref. \cite{Brascamp} and the method of
Ref. \cite{Ivasho}, in the present paper
we derive exact finite-size corrections for the free energy $F$ and
and the specific heat $C$ of the critical ferromagnetic Ising model on the
${\cal M} \times 2 {\cal N}$ square lattice with
Brascamp-Kunz (BK) boundary conditions \cite{Brascamp}
and compared such results with those under toroidal  boundary conditions
\cite{ff69,ih02}. We find that when the ratio $\xi/2=({\cal M}+1) / 2 {\cal N}$
is smaller than 1 the behaviors of finite-size corrections for $C$
are quite different for BK and toroidal boundary conditions;
when $\ln (\xi/2)$ is larger than 3, finite-size corrections for $C$
in two boundary conditions approach the same values.
In the limit  ${\cal N} \to \infty$ we obtain
the expansion of the free energy for infinitely long strip with
BK boundary conditions. Our results are consistent with the
conformal field theory prediction for the mixed boundary conditions
by Cardy \cite{Cardy1}
although the definitions of boundary conditions in two cases are
different in one side of the long strip.

This paper is organized as follows.
In Sec. II we show how to lead the partition function of Ising model
under Brascamp-Kunz (BK) boundary conditions to the form of partition
function with twisted boundary conditions. In Sec. III asymptotic
expansions of the free energy is presented.
In Sec. IV expansions of the specific heat is presented.
Our results are summarized and discussed in Sec. V.

\section{Ising model under Brascamp-Kunz\\ boundary conditions}
\label{First part}
For the Ising model on a lattice $G$ of $N$ sites, the $i$-th site of the
lattice for $1 \le i \le N$ is assigned a classical spin variable $s_{i}$,
which has values $\pm 1$.  The spins interact  according to the Hamiltonian
\begin{equation}
\beta H=-J\sum_{<ij>}{s_{i}s_{j}}
\label{hamilton}
\end{equation}
where $J$ is exchange energy, the sum runs over the nearest
neighbor pairs of spins, and $\beta=1/k_B T$ is the inverse
temperature.
The partition function of the Ising model is given by the sum over
all spin configurations on the lattice
\begin{equation}
Z_{Ising}(J)=\sum_{{s}}{e^{-\beta H(s)}}.
\label{Ising}
\end{equation}
 As is mentioned in {introduction} there are a few
boundary conditions for which the Ising model has been solved exactly.
Among  them is the special boundary conditions studied by Brascamp
and Kunz (BK) \cite{Brascamp}.
They considered a lattice with $2 {\cal N}$ sites
in the $x$ direction and ${\cal M}$ sites in the $y$ direction. The boundary
conditions are periodic in the $x$  direction; in the $y$ direction,
the spins are up (+1) along  the upper border of the resulting cylinder
and have the alternative  values along the lower border of the resulting
cylinder as is shown in  Fig. 1.
Recently Janke and Kenna \cite{Janke} has analyzed the Ising model in two
dimensions with these boundary conditions. They have derived exact
expressions for the finite-size scaling of the specific heat up to
the ${\cal M}^{-3}$ order. In this paper we obtain all exact finite-size
corrections for the free energy and
the specific heat. Moreover in our case the terms of
asymptotic expansion are analytical functions. They are related to
Kronecker's double series \cite{Weil} which in turn can be expressed by
elliptic $\theta$-functions \cite{Ivasho}.

For the BK boundary conditions, the Ising partition function given in
Ref. \cite{Brascamp} can be rewritten as
\begin{equation}
Z_{{\cal M},2 {\cal N}}=(\sqrt{2} e^{\mu})^{2 {\cal M} {\cal N}}
\prod_{i=1}^{{\cal N}} \prod_{j=1}^{\cal M} F(i,j)
\label{PartFunc}
\end{equation}
where  $\mu = 1/2 \ln{\sinh{2 J}}$ and
\begin{equation}
F(i,j)=4\left[2\sinh^2{\mu}+\sin^2{\left(\frac{\pi (i-1/2)}{2 {\cal N}}
\right)}+
\sin^2{\left(\frac{\pi j}{2 ({\cal M}+1)}\right)}\right].
\label{prodfunc}
\end{equation}
Now we try to express the partition function $Z_{{\cal M}, 2 {\cal N}}$
given by Eq. (\ref{PartFunc})
to the form of partition function with twisted boundary
conditions $Z_{\alpha,\beta}(\mu)$
\begin{equation}
Z^2_{\alpha,\beta}(\mu)=
\prod_{i=0}^{N-1}\prod_{j=0}^{M-1}4\left[\textstyle{
\;\sin^2\left(\frac{\pi (i+\alpha)}{N}\right)+
\sin^2\left(\frac{\pi (j+\beta)}{M}\right)+2\,{\rm sinh}^2\mu
\;}\right]
\label{Zab2}
\end{equation}
for which a general theory about
its asymptotic expansion has been given in Ref. \cite{Ivasho}.
For this purpose we can express the double products\\
$\prod_{i=0}^{2 {\cal N}-1}\prod_{j=0}^{2 {\cal M}+1}F(i+1,j)$ through
$\prod_{i=1}^{{\cal N}}\prod_{j=1}^{{\cal M}}F(i,j)$ as
\begin{equation}
\prod_{i=0}^{2 {\cal N}-1}\prod_{j=0}^{2 {\cal M}+1}F(i+1,j)=
\left(\prod_{i=1}^{{\cal N}}\prod_{j=1}^{{\cal M}}F(i,j)\right)^4
\prod_{i=0}^{2 {\cal N}-1}F(i+1,0)F(i+1,{\cal M}+1).
\label{kap}
\end{equation}
Here we use the properties of function $F(i,j)$
\begin{equation}
F(2 {\cal N}+1-k,j)=F(k,j) \quad \mbox{and} \quad F(i,2 {\cal M}+2-k)=F(i,k).
\label{hatkutjun}
\end{equation}
This transformation  leads the rectangular lattice
${\cal M} \times 2 {\cal N}$ under consideration to the
lattice  $2 ({\cal M}+1) \times 2 {\cal N}$.
In what follows we will use for convenience the
definition of the aspect ratio as
$\xi=\frac{{\cal M}+1}{{\cal N}}$ instead of conventional one
($\xi=\frac{{\cal M}}{2 {\cal N}}$).

The left-hand side of Eq. (\ref{kap}) is nothing but the
partition function with twisted boundary conditions $Z_{1/2,0}^2(\mu)$
given by Eq. (\ref{Zab2}) with $N = 2 {\cal N}$ and $M = 2 ({\cal M}+1)$.
With the help of the identity \cite{GradshteinRyzhik}
$$ \prod_{i=0}^{2 {\cal N}-1}4\textstyle{
\left[~\!{\rm sinh}^2\omega + \sin^2\left(\frac{\pi
(i+1/2)}{2 {\cal N}}\right)\right]} = 4 \cosh^2{(2 {\cal N}\omega)}
$$
the second product in the right-hand side of Eq. (\ref{kap}) can be
transformed into the form
\begin{equation}
\prod_{i=0}^{2 {\cal N}-1}F(i+1,0)F(i+1,{\cal M}+1) = \left[
4\cosh{\left(2 {\cal N}\omega_{\mu}(0)\right)}
\cosh{\left(2 {\cal N} \omega_{\mu}(\pi/2)
\right)}\right]^2
\label{prod}
\end{equation}
where
\begin{equation}
\omega_\mu(k)={\rm arcsinh}\sqrt{\sin^2 k+2\,{\rm sh}^2\mu}
\label{SpectralFunction}
\end{equation}
is a lattice dispersion relation.

By the aid of Eqs. (\ref{prodfunc}) - (\ref{kap}) and (\ref{prod}) the
partition function $Z_{{\cal M},2 {\cal N}}$ can be expressed as

\begin{equation}
Z_{{\cal M},2 {\cal N}}^{2}=\frac{\left(\sqrt{2}
e^{\mu}\right)^{4 {\cal M} {\cal N}}}
{4\cosh{\left[2 {\cal N} \omega_{\mu}(0)\right]}
\cosh{\left[2 {\cal N} \omega_{\mu}(\pi/2)
\right]}} \; Z_{1/2,0}(\mu).
\label{svjaz}
\end{equation}
For our further purposes we transform
the partition function $Z_{1/2,0}$   into the
simpler form
\begin{equation}
Z_{1/2,0}(\mu)=\prod_{n=0}^{N-1} 2 \textstyle{~\!{\rm
sinh}\left[M\omega_\mu\!\left(\frac{\pi(n+1/2)}{N}\right)
\right] }
\label{Zab}
\end{equation}
where  $N = 2 {\cal N}$ and $M = 2 ({\cal M}+1)$.
\section{Asymptotic Expansion of the Free Energy}
\label{Second part}
In the previous section it was shown that the partition function
of the ${\cal M} \times 2{\cal N}$ Ising model with
BK boundary conditions
can be expressed in terms of the partition
function with twisted boundary conditions $Z_{1/2,0}$, which has
been well studied in Ref. \cite{Ivasho}. Farther we
will use it and for simplicity we will remind some necessary
parts from there. For reader's convenience, all the technical
details of our calculations and definitions of the special
functions are summarized in the appendices at the end of this paper.
Considering the logarithm of the partition function with twisted
boundary conditions, Eq. (\ref{Zab}), we note, that it can be
transformed as
\begin{equation}
\ln Z_{1/2,0}(0)= M\sum_{n=0}^{N-1}
\omega_0\!\left(\textstyle{\frac{\pi(n+1/2)}{N}}\right)+
\sum_{n=0}^{N-1}\ln\left(\,1-e^{-2\,M
\omega_0\left(\frac{\pi(n+1/2)}{N}\right)}\right).
\label{lnZab}
\end{equation}
The second sum here vanishes in the limit $M \to \infty$ when
our lattice turns into infinitely long cylinder of circumference
$N$. Therefore, the first sum gives the logarithm of the partition
function with twisted angle $1/2$ on that cylinder. Its
asymptotic expansion can be found with the help of the
Euler-Maclaurin summation formula \cite{Hardy}

\begin{equation}
M\sum_{n=0}^{N-1}\omega_0\!\left(\textstyle{\frac{\pi(n+1/2)}{N}}\right)=
\frac{S}{\pi}\int_{0}^{\pi}\!\!\omega_0(x)~\!{\rm
d}x-\pi \xi\,{\rm B}_{2}^{1/2}- 2\pi\xi\sum_{p=1}^{\infty}
\left(\frac{\pi^2\xi}{S}\right)^{p}
\frac{\lambda_{2p}}{(2p)!}\;\frac{{\rm B}_{2p+2}^{1/2}}{2p+2}.
\label{EulerMaclaurinTerm}
\end{equation}
Here $S = N M =4 {\cal N}({\cal M}+1)$,
${\rm B}^{1/2}_{p}$ are the Bernoulli polynomials
${\rm B}^{\alpha}_{p}$ at $\alpha =1/2$ which are related
to the Bernoulli numbers $B_p \equiv B_p^0$ as $B_p^{1/2}=
(2^{1-p}-1)B_p$ and $\xi=\frac{M}{N}=\frac{{\cal M}+1}{{\cal N}}$.
We have also used the symmetry property,
$\omega_0(k)=\omega_0(\pi-k)$, of the lattice dispersion relation given by
Eq. (\ref{SpectralFunction}) and its Taylor expansion
\begin{equation}
\omega_0(k)=k\left(\lambda+\sum_{p=1}^{\infty}
\frac{\lambda_{2p}}{(2p)!}\;k^{2p}\right),
\label{SpectralFunctionExpansion}
\end{equation}
where $\lambda=1$, $\lambda_2=-2/3$, $\lambda_4=4$, etc.

We may transform the second term in Eq. (\ref{lnZab})  as

\begin{eqnarray}
&&\sum_{n=0}^{N-1}\ln\left(\,1-e^{-2\,M\omega_0\left(\frac{\pi(n+1/2)}
{N}\right)\,}\right)=
\ln\frac{\theta_{4}}{\eta}+\pi \xi\,{\rm
B}_2^{1/2}\nonumber\\[0.1cm]&&~~~~~~~~~~~~~~~~~-2\pi\xi\sum_{p=1}^{\infty}
\left(\frac{\pi^2\xi}{S}\right)^{p}\frac{\Lambda_{2p}}{(2p)!}
\,\frac{{\tt Re}\;\;{\rm K}_{2p+2}^{1/2,0}(i\lambda\xi)-{\rm
B}_{2p+2}^{1/2}}{2p+2} \label{ln(1-e)},
\end{eqnarray}
where $\eta=(\theta_2\theta_3\theta_4/2)^{1/3}$ is the
Dedekind-$\eta$ function; $\theta_2$, $\theta_3$, $\theta_4$ are elliptic
$\theta$-functions and ${\rm K}_{2p+2}^{1/2,0}(i\lambda\xi)$  are
Kronecker's double series \cite{Ivasho,Weil} (see also Appendix
\ref{KroneckerDoubleSeries}).
Taking into account the relation between moments and cumulants
(Appendix \ref{MomentsCumulants}), the differential operators $\Lambda_{2p}$
that have appeared here
can be expressed via coefficients $\lambda_{2p}$ of the expansion
of the lattice dispersion relation as
\begin{eqnarray}
{\Lambda}_{2}&=&\lambda_2, \nonumber\\
{\Lambda}_{4}&=&\lambda_4+3\lambda_2^2\,\frac{\partial}
{\partial\lambda}, \nonumber\\
{\Lambda}_{6}&=&\lambda_6+15\lambda_4\lambda_2\,\frac{\partial}
{\partial\lambda}
+15\lambda_2^3\,\frac{\partial^2}{\partial\lambda^2}, \nonumber\\
&\vdots&\nonumber\\ {\Lambda}_{p}&=&\sum_{r=1}^{p}\sum
\left(\frac{\lambda_{p_1}}{p_1!}\right)^{k_1}\ldots
\left(\frac{\lambda_{p_r}}{p_r!}\right)^{k_r}\frac{p!}{k_1!\ldots
k_r!}\;\frac{\partial^k}{\partial\lambda^k}. \nonumber
\end{eqnarray}
Here summation is over all positive numbers $\{k_1\ldots k_r\}$
and different positive numbers $\{p_1,\ldots,p_r\}$ such that $p_1
k_1+\ldots+ p_r k_r=p$ and $k=k_1+\ldots+k_r-1$.

Substituting Eqs. (\ref{EulerMaclaurinTerm}) and (\ref{ln(1-e)})
into Eq. (\ref{lnZab}) we finally obtain exact asymptotic expansion
of the logarithm of the partition function with twisted boundary
conditions in terms of the Kronecker's double series
\begin{eqnarray}
\ln
Z_{1/2,0}(0)&=&\frac{S}{\pi} \int_{0}^{\pi}\!\!\omega_0(x)~\!{\rm d}x  +
\ln\frac{\theta_4}{\eta}\nonumber\\[0.1cm]
&&-2\pi\xi\sum_{p=1}^{\infty}
\left(\frac{\pi^2\xi}{S}\right)^{p}\frac{\Lambda_{2p}}{(2p)!}\,
\frac{{\tt Re}\;{\rm K}_{2p+2}^{1/2,0}(i\lambda\xi)}{2p+2},
\label{ExpansionOflnZab}
\end{eqnarray}
where $\int_{0}^{\pi}\!\!\omega_0(x)~\!{\rm d}x = 2 G$
and $G = 0.915966$ is the Catalan's constant.

After reaching this point, one can easily write down the exact asymptotic
expansion of the free energy, $F=-\ln{Z_{{\cal M}, 2 {\cal N}}}$, at the
critical point. Plugging Eq. (\ref{ExpansionOflnZab}) back in
Eq. (\ref{svjaz}) we have finally obtain

\begin{eqnarray}
F &=& - 2 {\cal M} {\cal N} \left(\frac{1}{2}\ln{2}+\frac{2 G}{\pi}\right)
+2 {\cal N} \left[\frac{1}{2}\ln{(1+\sqrt{2})}
-\frac{2 G}{\pi}\right]
-\frac{1}{2}\ln{\frac{\theta_4}{2\eta}}
\nonumber\\
&+&
\pi\xi\sum_{p=1}^{\infty}
\left(\frac{\pi}{2{\cal N}}\right)^{2 p}\frac{\Lambda_{2p}}{(2p)!}\,
\frac{{\tt Re}\;{\rm K}_{2p+2}^{1/2,0}(i\lambda\xi)}{2p+2}.
\label{freeenergy}
\end{eqnarray}

Note that the Kroneker's functions $K_p^{\frac{1}{2},0}(i\lambda\xi)$
can be expressed in terms of the
elliptic $\theta$-function only. Thus, Eq. (\ref{freeenergy}) can be
rewritten in the following form
\begin{equation}
F = 2{\cal M}{\cal N} f_{bulk}+2{\cal N}f_{1}+f_0
+\sum_{p=1}^{\infty}\frac{f_{2p}}{\left({2{\cal N}}\right)^{2p}},
\label{freeenergy2}
\end{equation}
where
\begin{eqnarray}
f_{bulk}&=&-\frac{1}{2}\ln{2}-\frac{2 G}{\pi}=-0.929695...,
\label{fb}\\
f_{1}&=&\frac{1}{2}\ln{(1+\sqrt{2})}
-\frac{2 G}{\pi}=-0.142435...,
\label{f1}\\
f_0&=&-\frac{1}{2}\ln{\frac{\theta_4}{2\eta}},
\label{f0}\\
f_2&=&-\frac{\pi^3 \xi}{360}\left(\frac{7}{8} \theta_4^8
-\theta_2^4\theta_3^4\right),
\label{f2}\\
f_4&=&-\frac{\pi^5 \xi}{48384}\left[
\pi~ \xi~ \theta_3^4\theta_4^4\left(
\theta_3^8+\frac{5}{4}\theta_3^4\theta_4^4-\frac{5}{16}\theta_4^8\right)
\right.
\nonumber\\
&& \hspace{0.01cm}+\left.(\theta_2^4+\theta_3^4)
\left(\frac{31}{16}\theta_4^{8}+\theta_2^4\theta_3^4\right)
\left(1+4~\xi \frac{\theta_2'}{\theta_2}\right)\right],
\label{f4}\\
f_6&=&\frac{\pi^7 \xi}{87091200}
 \left[
70 \pi^2 \xi^2 \theta_3^4\theta_4^4
\left(\theta_2^{16}+
\frac{\theta_2^{12}\theta_4^4}{2}-\frac{81\theta_2^8\theta_4^{8}}{8}
\right. \right.
\nonumber\\
&&\hspace{6cm}\left. \left.
-\frac{295\theta_2^4\theta_4^{12}}{16}-\frac{635\theta_4^{16}}{64}\right)
\right.
\nonumber\\
&+&
\left.630 \pi \xi \theta_3^4\theta_4^4\left(\theta_2^{12}-
\frac{3\theta_2^8\theta_4^4}{4}-\frac{21\theta_2^4\theta_4^{8}}{8}
-\frac{127\theta_4^{12}}{64}\right)
\left(1+4~\xi \frac{\theta_2'}{\theta_2}\right)\right.
\nonumber\\
&+&\left.\left(\theta_2^{16}+2\theta_2^{12}\theta_4^4-
\frac{3}{4}\theta_2^8\theta_4^8-\frac{7}{4}\theta_2^4\theta_4^{12}
-\frac{127}{128}\theta_4^{16}\right)\right.
\label{f6}\\
&&\hspace{2.5cm}
\left.\left
(711+
5040\xi \frac{\theta_2'}{\theta_2}+8400\xi^2
\frac{\theta_2'^2}{\theta_2^2}
+560 \xi^2 \frac{\theta_2''}{\theta_2}
\right)\right],
\nonumber
\end{eqnarray}
\begin{eqnarray}
f_8&=&-\pi^{12}\xi^4\frac{\theta_4^4 \theta_3^4}{33634123776}
\left(1280\theta_2^{24}+20224\theta_2^{20}\theta_4^4+
83664\theta_2^{16}\theta_4^8\right.
\nonumber\\
&+&\left.210496\theta_2^{12}\theta_4^{12}+361115\theta_2^8
\theta_4^{16}+323910\theta_2^4\theta_4^{20}+107310\theta_4^{24}\right)
\nonumber\\
&-&\pi^{11}\xi^3\frac{\theta_4^4\theta_3^4\left(1+4\xi
\frac{\theta_2'}{\theta_2}\right)}{1868562432}\left(1280\theta_2^{20}
+8832\theta_2^{16}\theta_4^4+19056\theta_2^{12}\theta_4^8\right.
\nonumber\\
&+&\left. 33568\theta_2^8\theta_4^{12}+38655\theta_2^4\theta_4^{16}
+15330\theta_4^{20}\right)
\label{f8}\\
&-&\frac{\pi^{10}\xi^2\theta_4^4\theta_3^4}
{1177194332160}\left(3789+27720\xi\frac{\theta_2'}{\theta_2}+
48720\xi^2\frac{\theta_2'^2}{\theta_2^2}+2240\xi^2\frac{\theta_2''}
{\theta_2}\right)
\nonumber\\
&&\left(1280\theta_2^{16}+3136\theta_2^{12}\theta_4^4
+3216\theta_2^8\theta_4^8+5176\theta_2^4\theta_4{12}
+2555\theta_4^{16}\right)
\nonumber\\
&-&\frac{\pi^9\xi\left(\theta_2^4+\theta_3^4\right)}{235438866432}
\left(1479+15156\xi\frac{\theta_2'}{\theta_2}+47880\xi^3\frac{\theta_2'^2}
{\theta_2^2}+47880\xi^3\frac{\theta_2'^3}
{\theta_2^3}\right.
\nonumber\\
&+&\left.2520\xi^2\frac{\theta_2''}{\theta_2}+7980\xi^3
\frac{\theta_2'}{\theta_2}\frac{\theta_2''}{\theta_2}+
140\frac{\theta_2'''}{\theta_2}\right).
\nonumber\\
&\vdots&
\nonumber
\end{eqnarray}
The free energy per unit length of an infinitely long strip of width $L$
at critically has the finite-size scaling form \cite{Blote}
\begin{equation}
F=fL+f^*+\frac{\Delta}{L}+\ldots,
\label{strip}
\end{equation}
where $f$ is the bulk free energy per unit area,  $\frac{1}{2}f^*$
is the surface energy, $L^{-1}$ is a scaling field, and $\Delta$ is an
universal constant which depends only on the type of boundary conditions
\cite{Cardy1},
\begin{eqnarray}
\Delta&=& -\frac{\pi}{12} \qquad \mbox{periodic boundary conditions}
\nonumber, \\
\Delta&=& \frac{\pi}{6} \qquad \quad \mbox{antiperiodic boundary conditions}
   \nonumber, \\
\Delta&=& -\frac{\pi}{48} \qquad \mbox{free boundary conditions}\nonumber, \\
\Delta&=& -\frac{\pi}{48} \quad \quad \mbox{fixed $++$ boundary conditions}
\nonumber, \\
\Delta&=& \frac{23\pi}{48} \quad \quad \mbox{fixed $+-$ boundary conditions}
\nonumber, \\
\Delta&=& \frac{\pi}{24} \qquad \quad \mbox{mixed boundary conditions}.
\label{Delta}
\end{eqnarray}
For fixed $++$ (or $+-$) boundary conditions the spins are fixed to the same
(or opposite) values on two sides of the strip. The mixed boundary
conditions corresponds to free boundary conditions on one side of the strip,
and fixed boundary conditions on the other. Therefore, BK and the mixed
boundary
conditions are the same in one side of the long strip
(fixed to + for all spins) and
they are different in another side of the long strip (fixed to $+-+-...$
for BK boundary conditions and free boundary conditions for the mixed
boundary conditions).

Using Kronecker's functions asymptotic form when
$\xi \to 0$ and $\xi \to \infty$
 we can obtain from Eq. (\ref{freeenergy}) the free energy per unit
length of an infinitely long strip of finite width.
In the limit $\xi \to \infty$ (i.e. ${\cal M} \to \infty$) for fixed
$2 {\cal N}$ from Eq. (\ref{freeenergy}) one obtains the free energy
expansion for
infinitely long cylinder of circumference $2 {\cal N}$

\begin{eqnarray}
\lim_{{\cal M} \to \infty}\frac{F}{{\cal M}}&=&2{\cal N} f_{bulk}
-\frac{\pi}{24 {\cal N}}
+2\sum_{p=1}^{\infty}\left(\frac{\pi}{2{\cal N}}\right)^{2p+1}
\frac{\lambda_{2p}B_{2p+2}^{1/2}}{(2p)!(2p+2)}
\label{Blotef2N}\\
&=&2 {\cal N} \left(-\frac{2 G}{\pi}-
\frac{1}{2}\ln{2}\right)-
\frac{\pi}{12}\left(\frac{1}{2{\cal N}}\right)-
\frac{7\pi^3}{1440}{\left(\frac{1}{2{\cal N}}\right)}^3
\nonumber\\
&-&
\frac{31\pi^5}{24192}\left(\frac{1}{2{\cal N}}\right)^5
-\frac{10033\pi^7}{9676800}{\left(\frac{1}{2{\cal N}}\right)}^7 -\ldots.
\nonumber
\end{eqnarray}
This result coincides with that obtained in \cite{ih01}
with the leading finite-size correction to free energy
$-\frac{\pi}{12}(2{\cal N})^{-1}$.
In the limit $ \xi \to 0$ (i.e. ${\cal N} \to \infty$) for fixed ${\cal M}$
we obtain the expansion of free energy of infinitely long strip
with BK boundary condition of the width ${\cal M}$
\begin{eqnarray}
\lim_{{\cal N} \to \infty}\frac{F}{2{\cal N}}&=&{\cal M} f_{bulk}+f_1
+\frac{\pi}{24({\cal M}+1)}
\nonumber\\
&+&\sum_{p=1}^{\infty}\left[\frac{\pi}{2({\cal M}+1)}\right]^{2p+1}
\frac{\lambda_{2p}B_{2p+2}}{(2p)!(2p+2)}
\label{BlotefM}\\
&=&{\cal M}\left(-\frac{2 G}{\pi}-
\frac{1}{2}\ln{2}\right) + \frac{1}{2}\ln{(1+\sqrt{2})}
-\frac{2 G}{\pi}
+\frac{\pi}{24}\frac{1}{{\cal M}}
\nonumber\\
&-&\frac{\pi}{24}\frac{1}{{\cal M}^2}
+\left(\frac{\pi}{24}+\frac{\pi^3}{2800}\right)\frac{1}{{\cal M}^3}
- \left(\frac{\pi}{24}+\frac{\pi^3}{960}\right)\frac{1}{{\cal M}^4}+
\ldots.
\nonumber
\end{eqnarray}
Here the leading finite-width correction to free
energy is $\frac{\pi}{24}{\cal M}^{-1}$.
>From Eqs. (\ref{strip}) - (\ref{BlotefM}) one can see, that
\begin{eqnarray}
&1.&\;\;\;L=2{\cal N},\;\;f=f_{bulk},\;\;f^*=0,\;\;\Delta=-\frac{\pi}{12},
\label{depq1}\\
&2.&\;\;\;L={\cal M},\;\;f=f_{bulk},\;\;f^*=f_1,\;\;\Delta=\frac{\pi}{24}.
\label{depq2}
\end{eqnarray}
Our results are consistent with the
conformal field theory prediction for the
mixed boundary condition (see Eq. (\ref{Delta}))
although the mixed boundary condition and
the BK boundary condition are different in one side of the long strip.

\section{Asymptotic Expansion of the Internal\\ Energy and
the Specific Heat}
\label{subsec:2a}
The internal energy per spin and the specific heat per spin
 can be obtained from the partition function $Z_{{\cal M},2{{\cal N}}}$
\begin{eqnarray}
U&=&-\frac{1}{2 {\cal M} {\cal N}} \frac{d}{d J}\ln{Z_{{\cal M},2 {\cal N}}}
=-\frac{\sqrt{1+e^{-4\mu}}}{2 {\cal M} {\cal N}}
\frac{d}{d \mu}\ln{Z_{{\cal M},2{\cal N}}},
\label{internalenergy1}
\end{eqnarray}
\begin{eqnarray}
C&=& \frac{1}{2 {\cal M} {\cal N}} \frac{d^2}{d J^2}\ln{Z_{{\cal M},2
{\cal N}}}\nonumber \\
 &=& \frac{e^{-4\mu}}{{\cal M} {\cal N}}
\left(\frac{1+e^{4\mu}}{2}\frac{d^2}{d {\mu}^2}\ln{Z_{{\cal M},2 {\cal N}}}
-\frac{d}{d \mu}\ln{Z_{{\cal M},2 {\cal N}}}\right).
\label{spheat1}
\end{eqnarray}
Let us first consider the internal energy.
At the critical point $T=T_c$ ($\mu = 0$) the internal energy is given by
\begin{equation}
U=-\sqrt{2} +\sqrt{2}\frac{d}{d \mu}\ln Z_{1/2,0}(0).
\label{intenergy}
\end{equation}
One can note that $Z_{1/2,0}(\mu)$ is an even function with respect to its
argument $\mu$, which imply immediately that
$\left(d Z_{1/2,0}(\mu)/d\mu\right)_{\mu=0}=0$. Thus we find that
internal energy for the finite system is equal to its bulk values without
any finite-size corrections, namely $U=-\sqrt{2}$.

At the critical point $T=T_c$ ($\mu = 0$) the specific heat is given by
\begin{equation}
C=-2 -\frac{4 {\cal N}}{{\cal M}}-\frac{\sqrt{2}}{{\cal M}}
\tanh{\left[2 {\cal N}\ln{(1+\sqrt{2})}\right]}+\frac{1}{2 {\cal M} {\cal N}}
\frac{d^2}{d \mu^2}
\ln Z_{1/2,0}(0).
\label{intenergyspheat}
\end{equation}
The analysis of the $Z''_{1/2,0}(0)$ is a little more involved.
Taking the second derivative of
Eq.(\ref{Zab}) with respect to mass variable $\mu$ and then
considering the limit $\mu\to 0$, we obtain

\begin{eqnarray}
\frac{Z''_{1/2,0}(0)}{Z_{1/2,0}(0)}&=&{\tt
}M \sum_{n=0}^{N-1}
\omega''_0\!\left(\textstyle{\frac{\pi(n+1/2)}{N}}\right) {\rm
cth}\left[M\omega_0\!\left(\textstyle{\frac{\pi(n+1/2)}{N}}\right)
\right]
\label{ExpansionZ''} \\
&=&M \sum_{n=0}^{N-1}\omega''_0\!\left(\textstyle{\frac{\pi(n+1/2)}{N}}
\right)+2~{\tt } M \sum_{n=0}^{N-1}\sum_{m=1}^{\infty}
\omega''_0\!\left(\textstyle{\frac{\pi(n+1/2)}{N}}\right)
e^{-2m\left[M\omega_0\left(\frac{\pi(n + 1/2)}{N}\right)
\right]},
\nonumber
\end{eqnarray}
where $M = 2({\cal M}+1)$, $N = 2 {\cal N}$, and $\omega''_0(x)$ is the
second derivative of $\omega_{\mu}(x)$ with respect to $\mu$ at criticality
\begin{eqnarray}
\omega''_0(x) = \frac{2}{\sin{x}\;\sqrt{1+\sin^2{x}}}.
\nonumber
\end{eqnarray}
Using Taylor's theorem, the asymptotic expansion of the  $\omega''_0(x)$  can
be written in the following form
\begin{eqnarray}
\omega''_0(x) = \frac{2}{x}\left\{1+\sum_{p=1}^{\infty}\frac
{\kappa_{2p}}{(2p)!}x^{2p}
\right\},
\nonumber
\end{eqnarray}
where $\kappa_{2}=-2/3$, $\kappa_{4}=172/15$, etc.
The first sum in Eq. (\ref{ExpansionZ''}) we may transform as
\begin{equation}
M \sum_{n=0}^{N-1}\omega''_0\!\left(\textstyle{\frac{\pi(n+1/2)}{N}}
\right) = M \sum_{n=0}^{N-1} f\!\left(\textstyle{\frac{\pi(n+1/2)}{N}}
\right)
+ \frac{4 S}{\pi} \sum_{n=0}^{N-1}\frac{1}{n+1/2},
\label{newsum}
\end{equation}
where we have introduce the function $f(x)=\omega''_0(x)-2/x-
2/(\pi - x)$.
This function and all its derivatives are integrable over the interval
$(0,\pi)$. Thus, for the first term in Eq. (\ref{newsum}) we may use again
the Euler-Maclaurin summation formula,
and after a little algebra we obtain
\begin{eqnarray}
M \sum_{n=0}^{N-1} f\!\left(\textstyle{\frac{\pi(n+1/2)}{N}}
\right)&=&\frac{S}{\pi}\int_{0}^{\pi}\!\!f(x)~\!{\rm
d}x - 2 \pi\xi \sum_{p=1}^{\infty}
\left(\frac{\pi^2\xi}{S}\right)^{p-1}
\frac{\kappa_{2p}{\rm B}_{2p}^{1/2}}{p(2p)!}
\nonumber\\
&+&\frac{2 S}{\pi}\sum_{p=1}^{\infty}
\frac{{\rm B}_{2p}^{1/2}}{p}\frac{1}{N^{2p}},
\label{ExpansionF}
\end{eqnarray}
where $\int_{0}^{\pi}\!\!f(x)~\!{\rm d}x=2\ln{2}-4\ln{\pi}$.
The second sum in Eq.  (\ref{newsum}) can be written in terms of the
digamma function $\psi(x)$.
\begin{equation}
\sum_{n=0}^{N-1} \frac{1}{n+1/2} = \psi(N+1/2)-\psi(1/2).
\label{psi}
\end{equation}
The asymptotic expansion of the digamma function $\psi(x)$ is given by
(see Appendix \ref{PsiSeries})
\begin{equation}
\psi(N+1/2)=\ln{N}-\sum_{p=1}^{\infty}(-1)^p
\frac{{\rm B}_{p}^{1/2}}{p}\frac{1}{N^p}.
\label{ExpansionPsi}
\end{equation}
Using the property of the Bernoulli polynomials
${\rm B}_{p}^{1/2}$, namely, ${\rm B}_{2p+1}^{1/2}=0$, Eq. (\ref{psi})
can be rewritten as
\begin{equation}
\sum_{n=0}^{N-1} \frac{1}{n+1/2}
= \ln{N}-
\sum_{p=1}^{\infty}
\frac{{\rm B}_{2p}^{1/2}}{2 p}\frac{1}{N^{2p}}-\psi(1/2).
\label{psi1}
\end{equation}
Plugging Eqs. (\ref{ExpansionF}) and (\ref{psi1}) back in Eq. (\ref{newsum}),
we have finally obtain
\begin{eqnarray}
M \sum_{n=0}^{N-1}\omega''_0\!\left(\textstyle{\frac{\pi(n+1/2)}{N}}
\right) &=&\frac{4 S}{\pi}\left\{\ln{N}+\frac{1}{2}\ln{2}-\ln{\pi}
-\psi(1/2)\right\}
\nonumber\\
&-& 2 \pi\xi \sum_{p=1}^{\infty}
\left(\frac{\pi^2\xi}{S}\right)^{p-1}
\frac{\kappa_{2p}{\rm B}_{2p}^{1/2}}{p(2p)!}.
\label{newsum1}
\end{eqnarray}

Let us now consider the second sum in Eq. (\ref{ExpansionZ''}). Note that
function $\omega''_0(x)$ can be represented as
\begin{equation}
\omega''_0(x)=\frac{2}{x}\;\exp\left\{
{\sum_{p=1}^{\infty}\frac{\varepsilon_{2p}}{(2p)!}}x^{2p}
\right\},
\label{owega''}
\end{equation}
where coefficients $\varepsilon_{2p}$ and $\kappa_{2p}$ are related to each
other through relation between moments and cumulants
(Appendix \ref{MomentsCumulants}). Following along the same lines as in the
section (3), the second sum
in Eq. (\ref{ExpansionZ''}) can be written as

\begin{eqnarray}
&&2~{\tt } M \sum_{n=0}^{N-1}\sum_{m=1}^{\infty}
\omega''_0\!\left(\textstyle{\frac{\pi(n+1/2)}{N}}\right)
e^{-2mM\omega_0\left(\frac{\pi(n + 1/2)}{N}\right)}=
\nonumber\\
&&=\frac{4 S}{\pi}\left\{R_{1/2,0}(\xi)
+\psi(1/2)\right\} +\left(\kappa_2\xi\frac{\partial}{\partial\xi}+
\lambda_2\xi^2
\frac{\partial^2}{\partial\xi^2}\right)
\ln\frac{\theta_{4}(\xi)}{\eta(\xi)}
\nonumber\\
&&-2 \pi \xi \sum_{p=2}^{\infty}\frac{\Omega_{2p}}{p(2p)!}
\left(\frac{\pi^2 \xi}{S}\right)^{p-1}{\tt Re}\;
{\rm K}_{2p}^{1/2,0}(i\lambda\xi)+
2 \pi \xi \sum_{p=1}^{\infty}\frac{\kappa_{2p}
\;{\rm B}_{2p}^{1/2}}{p(2p)!}
\left(\frac{\pi^2 \xi}{S}\right)^{p-1}, \label{ExpansionOmega''2}
\end{eqnarray}
where
\begin{eqnarray}
R_{1/2,0}(\xi) =
-2 \ln{\theta_{4}(\xi)} + C_E + 2\ln{2}
\nonumber
\end{eqnarray}
and $C_E$ is the Euler constant.
The differential operators $\Omega_{2p}$ that have appeared here
can be expressed via coefficients $\omega_{2p}=\varepsilon_{2p}+\lambda_{2p}
\frac{\partial}{\partial\lambda}$ as
\begin{eqnarray}
{\Omega}_{2}&=&\omega_2, \nonumber\\
{\Omega}_{4}&=&\omega_4+3\omega_2^2,
\nonumber\\
&\vdots&
\nonumber
\end{eqnarray}
Substituting Eqs. (\ref{newsum1}) and (\ref{ExpansionOmega''2}) into
Eq. (\ref{ExpansionZ''}), we obtain exact
asymptotic expansion of $Z''_{1/2,0}(0)$
\begin{eqnarray}
\frac{Z''_{1/2,0}(0)}{Z_{1/2,0}(0)}
&=&\frac{4 S}{\pi}\left(\ln{N}+C_E+\ln{\frac{2^{5/2}}{\pi}} -
2 \ln{\theta_{4}(\xi)}
\right)
\nonumber\\
&+&\left(\kappa_2\xi\frac{\partial}{\partial\xi}+\lambda_2\xi^2
\frac{\partial^2}{\partial\xi^2}\right)
\ln\frac{\theta_{4}(\xi)}{\eta(\xi)}
\label{ExpansionZ''fin}\\
&-& 2 \pi \xi \sum_{p=2}^{\infty}\frac{\Omega_{2p}}{p(2p)!}
\left(\frac{\pi^2 \xi}{S}\right)^{p-1}{\tt Re}\;
{\rm K}_{2p}^{1/2,0}(i\lambda\xi).
\nonumber
\end{eqnarray}
Plugging  Eq. (\ref{ExpansionZ''fin}) back in Eq. (\ref{intenergyspheat})
we finally obtain exact asymptotic expansion of the specific heat
\begin{equation}
C=\frac{8}{\pi}\left(1+\frac{1}{\xi {\cal N}-1}\right)\ln{2 {\cal N}}
+\sum_{p=0}^{\infty}\frac{C_{p}}{(2{\cal N})^{p}},
\label{1/Nexp}
\end{equation}
where
\begin{eqnarray}
C_0&=&\frac{8}{\pi}\left(C_E
+\ln{\frac{2^{5/2}}{\pi}}-\frac{\pi}{4}\right)-\frac{4}{\xi}
 -\frac{16}{\pi} \ln{\theta_{4}(\xi)},
\nonumber\\
C_1&=&\frac{2}{\xi}(C_0+2-\sqrt{2}),
\nonumber\\
C_2&=&\frac{2}{\xi}C_1-\frac{\pi}{9}\left\{\pi\xi \theta_3^4\theta_4^4+
(\theta_2^4+\theta_3^4)\left(1+4\xi\frac{\theta_2'}{\theta_2}\right)\right\},
\nonumber\\
C_3&=&\frac{2}{\xi}C_2,
\nonumber\\
C_4&=&\frac{2}{\xi} C_3
+\frac{\pi^5 \xi^4 \theta_4^4 \theta_3^4}{270}
\left(\theta_2^8-\frac{3}{2}\theta_4^4\theta_2^4-\frac{21}{4}\theta_4^8\right)
\nonumber\\
&+&\frac{\pi^4 \xi^3 \theta_4^4 \theta_3^4 }{54}
\left(\theta_2^4-\frac{7}{4}\theta_4^4\right)\left(1+
4\xi\frac{\theta_2'}{\theta_2}\right)
\nonumber\\
&+&\frac{4\pi^3 \xi^2}{135}\left(\theta_2^4\theta_3^4-\frac{7}{8}\theta_4^8
\right)\left(\frac{43}{40}+5\xi\frac{\theta_2'}{\theta_2}+
7\xi^2\frac{\theta_2'^2}{\theta_2^2}
+\xi^2\frac{\theta_2''}{\theta_2}\right),
\nonumber\\
C_5&=&\frac{2}{\xi}C_4,
\nonumber\\
&\vdots&
\nonumber
\end{eqnarray}

The $\frac{1}{{\cal M}}$ expansion of the specific heat  has a form

\begin{equation}
C=\frac{8}{\pi}\left(1+\frac{1}{{\cal M}}\right)\ln{{\cal M}}
+\sum_{p=0}^{\infty}\frac{c_{p}}{{\cal M}^p},
\label{1/Mexp}
\end{equation}
where
\begin{eqnarray}
c_0&=&C_0 -\frac{8}{\pi} \ln{\frac{\xi}{2}},
\nonumber\\
c_1&=&\frac{\xi}{2}C_1+\frac{8}{\pi}(1-\ln{\frac{\xi}{2}}),
\nonumber\\
c_2&=&\frac{\xi^2}{4} C_2- \frac{\xi}{2} C_1+
\frac{4}{\pi},
\nonumber\\
c_3&=&-\frac{\xi^2}{4} C_2+\frac{\xi}{2} C_1-
\frac{4}{3\pi},
\nonumber\\
c_4&=&\frac{\xi^4}{16}C_4-\frac{\xi}{2} C_1+\frac{2}{3\pi},
\nonumber \\
c_5&=&-\frac{3\xi^4}{16}C_4+\frac{\xi^2}{2} C_2+\frac{\xi}{2} C_1-
\frac{2}{5\pi},
\nonumber \\
\vdots
\nonumber
\end{eqnarray}
Typical value of the constants $c_{0}$ - $c_{3}$ are given in
Table 1, in which the coefficients are consistent with those
 obtained in \cite{Janke}.

Using Kronecker's functions asymptotic form when $\xi \to 0$ and
$\xi \to \infty$ we can obtain from Eqs. (\ref{intenergyspheat}) and
(\ref{ExpansionZ''fin}) the specific heat per unit length of an infinitely
long strip of finite width.
In the limit $\xi \to \infty$ (i.e. ${\cal M} \to \infty$) for fixed
$2{\cal N}$  the specific heat expansion for
infinitely long cylinder of circumference $2 {\cal N}$
 can be written as
\begin{eqnarray}
C&=& \frac{8}{\pi}\ln{2{\cal N}}+\frac{8}{\pi}\left(C_E
+\ln{\frac{2^{5/2}}{\pi}}-\frac{\pi}{4}\right)-\sum_{p=1}^{\infty}
\frac{4 \pi^{2p-1}\Omega_{2p}B_{2p}^{1/2}}{p(2p)!}
\frac{1}{(2{\cal N})^{2p}}
\nonumber\\
&=&\frac{8}{\pi}\ln{2{\cal N}}+\frac{8}{\pi}\left(C_E
+\ln{\frac{2^{5/2}}{\pi}}-\frac{\pi}{4}\right)
-\frac{\pi}{9}\left(\frac{1}{2{\cal N}}\right)^2
\label{1/Nspheat}\\
&-&\frac{301\pi^3}{10800}\left(\frac{1}{2{\cal N}}\right)^4
-\frac{29419\pi^5}{1905120}\left(\frac{1}{2{\cal N}}\right)^6-
\frac{2759329\pi^7}{145152000}\left(\frac{1}{2{\cal N}}\right)^8
-\cdots.
\nonumber
\end{eqnarray}
In the limit $\xi \to 0$ (i.e. ${\cal N} \to \infty$) for fixed ${\cal M}$
we obtain the expression of specific heat of infinitely long strip with
BK boundary condition of width ${\cal M}$
\begin{eqnarray}
C&=&\frac{8}{\pi}\frac{{\cal M}+1}{\cal M}
\left(\ln{({\cal M}+1)}+C_E+\ln{\frac{2^{3/2}}{\pi}}\right)-2-
\frac{\sqrt{2}}{{\cal M}}
\nonumber\\
&-&\sum_{p=1}^{\infty}\frac{2 \, k_{2p}B_{2p}}
{p(2p)!}
\left(\frac{\pi}{2}\right)^{2 p-1}\frac{1}{{\cal M}({\cal M}+1)^{2p-1}}
\nonumber\\
&=&\frac{8}{\pi}\left(1+\frac{1}{\cal M}\right)\ln{{\cal M}}
+\frac{8}{\pi}
\left(C_E+\ln{\frac{2^{3/2}}{\pi}}-\frac{\pi}{4}\right)
\label{1/Mspheat}\\
&+&\frac{8}{\pi}
\left(C_E+\ln{\frac{2^{3/2}}{\pi}+1-\frac{\sqrt{2}\pi}{8}}\right)
\frac{1}{{\cal M}}+
\left(\frac{4}{\pi}+\frac{\pi}{18}\right)\left(\frac{1}{{\cal M}}\right)^2
\nonumber\\
&-&\left(\frac{4}{3\pi}+\frac{\pi}{18}\right)\left(\frac{1}{{\cal M}}
\right)^3
+\left(\frac{2}{3\pi}+\frac{\pi}{18}+\frac{43\pi^3}{21600}\right)
\left(\frac{1}{{\cal M}}\right)^4
\nonumber\\
&-& \left(\frac{2}{5\pi}+\frac{\pi}{18}+\frac{43\pi^3}{7200}\right)
\left(\frac{1}{{\cal M}}\right)^5
+\cdots.
\nonumber
\end{eqnarray}
Note that the specific heat expansion for
infinitely long cylinder  contains only even
powers of ${\cal N}^{-1}$ (except of course the leading logarithmic term),
while in the specific heat  expansion for infinitely long strip with
BK boundary condition  any integer powers
of ${\cal M}^{-1}$ can  occur.

In Fig. 2 we plot the aspect-ratio $(\xi)$ dependence of the finite-size
specific heat correction terms $C_0, C_1, C_2,$ and $C_3$ for the
Ising model with BK boundary condition and those of the
torus \cite{ih02}. We use the
logarithmic scales for the horizontal axis.
 For large enough
$\xi$ ($\gg 1$), the finite size properties of the Ising model with
BK boundary condition and those of the torus become the same
because the boundaries along the shorter direction determine the
finite size properties of the system; for both BK boundary
condition and the torus, the boundary condition along the $y$ axis is the
periodic one.

\section{Summary and discussion}
\label{sec:4}

In this paper, we have used the method of \cite{Ivasho} to derive
exact finite-size corrections for the free energy $F$
and the specific heat $C$ of the critical ferromagnetic Ising model on the
${\cal M} \times 2 {\cal N}$ square lattice with Brascamp-Kunz (BK) boundary
conditions \cite{Brascamp}. We find that the finite-size corrections to the
free energy and the specific heat are always integer powers of
${\cal N}^{-1}$ (${\cal M}^{-1}$) except of course the leading logarithmic
term in the  specific heat. In the finite-size expansion of the
free energy given by  Eq. (\ref{freeenergy2}), only even power of
${\cal N}^{-1}$ occur, except for the term ${\cal N}$.
In the finite-size expansion of the specific heat given by Eqs. (\ref{1/Nexp})
and (\ref{1/Mexp}),
any integer powers of
${\cal N}^{-1}$ (${\cal M}^{-1}$) can occur.

We have compared our results with those under toroidal  boundary conditions.
 When the ratio $\xi/2=({\cal M}+1) / 2 {\cal N}$
is smaller than 1 the behaviors of finite-size corrections for $C$
are quite different for BK and toroidal boundary conditions;
when $\ln (\xi/2)$ is larger than 3, finite-size corrections for $C$
in two boundary conditions approach the same values.
In the limit  ${\cal N} \to \infty$ we obtain
the expansion of the free energy for infinitely long strip with
BK boundary conditions. Our results are consistent with the
conformal field theory prediction for the mixed boundary conditions
by Cardy \cite{Cardy1}
although the definitions of boundary conditions in two cases are
different in one side of the long strip. It is of interest to know
under what conditions different boundary conditions could still give
the same finite-size corrections.

The results of this paper shows that the method of Ref. \cite{Ivasho}
is quite useful for calculating exact finite-size corrections for
critical systems. It is of interest to apply this method to calculate
exact finite-size corrections for the Ising model and other free models
\cite{Ivasho} on various lattices with various boundary conditions
so that some general features of such finite-size corrections
could be found.

This work was supported in part by the National Science
Council of the Republic of China (Taiwan) under Grant No.
NSC 90-2112-M-001-074.

\newpage
\centerline{\bf APPENDIX}
\appendix


\section{Kronecker's Double Series}
\label{KroneckerDoubleSeries}

Kronecker's double series can be defined as \cite{Weil}
$${\rm K}_{p}^{1/2,0}(\tau)= -\frac{p!}{(-2\pi i)^p}
\sum_{m,n\in Z \above0pt  (m,n)\neq(0,0)} \frac{e^{-\pi
in}}{(n+\tau m)^{p}}.$$
In this form, however, they cannot be directly applied to our
analysis. We need to cast them in a different form. To this end,
let us separate from the double series a subseries with $m=0$
$${\rm K}_{p}^{1/2,0}(\tau)=-\frac{p!}{(-2\pi i)^p}
\sum_{n\neq 0}\frac{e^{-\pi in}}{n^{p}} -\frac{p!}{(-2\pi
i)^p} \sum_{m\neq 0}\sum_{n\in Z} \frac{e^{-\pi
in}}{(n+\tau m)^{p}}.$$
Here the first sum gives nothing but Fourier representation of
Bernoulli polynomials
\begin{equation}
{\rm B}^{\alpha}_p=-\frac{p!}{(-2\pi i)^p}\sum_{n\neq
0}\frac{e^{-2\pi in \alpha}}{n^{p}}. \label{FourierBernoulli}
\end{equation}
The second sum
can be rearranged with the help of the identity
$$\frac{p!}{(-2\pi i)^p}\sum_{n\in Z}\frac{e^{-\pi
in}}{(z+n)^p}= p\sum_{n=0}^{\infty}(n+1/2)^{p-1}e^{2\pi i
z(n+1/2)},$$
which can easily be derived from  following equation
\begin{equation}
\frac{e^{2\pi iz\alpha}}{e^{2\pi
iz}-1}=-\sum_{n=0}^{\infty}e^{2\pi iz(n+\alpha)}=\frac{1}{2\pi i}
\sum_{n=-\infty}^{+\infty}\frac{e^{-2\pi i n \alpha}}{z + n}
\label{FourierGeneratingFunction}
\end{equation}
by differentiating it $p$ times.
The final result of our resummation of double Kronecker sum is
$${\rm K}_{p}^{1/2,0}(\tau)={\rm B}_{p}^{1/2}-p\sum_{m\neq
0}\sum_{n=0}^{\infty} (n+1/2)^{p-1}~e^{2\pi im
\tau(n+1/2)}.$$
Considering the Kronecker sums with pure imaginary
aspect ratio, $\tau=i\xi$, we can further rearrange this expression
to get summation only over positive $m\geq 1$
\begin{equation}
{\rm B}_{2p}^{1/2}-{\tt }~\!{\rm
K}_{2p}^{1/2,0}(i\xi)=4p\sum_{m=1}^{\infty}
\sum_{n=0}^{\infty} (n+1/2)^{2p-1}~e^{-2\pi m
\xi(n+1/2)}.
\label{IdentityKronecker}
\end{equation}
\section{Relation between moments and cumulants}
\label{MomentsCumulants}

Moments $Z_{2k}$ and cumulants $F_{2k}$ which enters the expansion of
exponent
$$\exp\left\{\;\sum_{k=1}^{\infty}\frac{x^{2k}}{(2k)!}\,F_{2k}\,\right\}
=1+\sum_{k=1}^{\infty}\frac{x^{2k}}{(2k)!}\,Z_{2k}$$
are related to each other as \cite{Prohorov}
\begin{eqnarray}
Z_2&=&F_2
\nonumber,\\
Z_4&=&F_4+3 F_2^2
\nonumber,\\
Z_6&=&F_6+15 F_2 F_4+15 F_2^3
\nonumber,\\
Z_8&=&F_8+28 F_2 F_6+35 F_4^2+210 F_2^2 F_4+105 F_2^4
\nonumber,\\
~&\vdots&
~\nonumber\\
Z_{k}&=&\sum_{r=1}^{{k}}\sum
\left(\frac{F_{k_1}}{k_1!}\right)^{i_1}\ldots
\left(\frac{F_{k_r}}{k_r!}\right)^{i_r} \frac{k!}{i_1!\ldots
i_r!},\nonumber
\end{eqnarray}
where summation is over all positive numbers $\{i_1\ldots i_r\}$
and different positive numbers $\{k_1,\ldots,k_r\}$ such that $k_1
i_1+\ldots+ k_r i_r=k$.

\section{Reduction of Kronecker's Double Series to Theta Functions}
\label{KroneckerToTheta}

Let us consider Laurent expansion of the Weierstrass function
\begin{eqnarray}
\wp(z)&=&\frac{1}{z^2}+\sum_{(n,m)\neq(0,0)}\left[\frac{1}{(z-n-\tau
m)^2}-\frac{1}{(n+\tau m)^2}\right]\nonumber\\
&=&\frac{1}{z^2}+\sum_{p=2}^{\infty}a_{p}(\tau) z^{2p-2}.\nonumber
\end{eqnarray}
The coefficients $a_p(\tau)$ of the expansion can all be written
in terms of the elliptic $\theta$-functions with the help of the
recursion relation \cite{Korn}
$$a_p=\frac{3}{(p-3)(2p+1)}~
(a_{2}a_{p-2}+a_{3}a_{p-3}+\ldots+a_{p-2}a_{2}),$$
where first terms of the sequence are
\begin{eqnarray}
a_2&=&\textstyle{\frac{\pi^4}{15}}(\theta_2^4\theta_3^4-
\theta_2^4\theta_4^4+\theta_3^4\theta_4^4)\nonumber,\\
a_3&=&\textstyle{\frac{\pi^6}{189}}(\theta_2^4+\theta_3^4)
(\theta_4^4-\theta_2^4)(\theta_3^4+\theta_4^4)\nonumber,\\
a_4&=&\textstyle{\frac{1}{3}}a_2^2\nonumber,\\
a_5&=&\textstyle{\frac{3}{11}}(a_2a_3)\nonumber,\\
a_6&=&\textstyle{\frac{1}{39}}(2a_2^3+3a_3^2)\nonumber,\\
&\vdots&\nonumber
\end{eqnarray}
Kronecker functions ${\rm K}_{2p}^{0,0}(\tau)$ are related
directly to the coefficients $a_p(\tau)$
$${\rm K}^{0,0}_{2p}(\tau)=
-\frac{(2p)!}{(-4\pi^2)^p}\frac{a_p(\tau)}{(2p-1)}.$$
Kronecker functions ${\rm K}_{2p}^{1/2,0}(\tau)$
 can in their turn
be related to the function ${\rm K}_{2p}^{0,0}(\tau)$ by means of
simple resummation of Kronecker's double series
\begin{eqnarray}
{\rm K}_{p}^{\frac{1}{2},0}(\tau)=2^{1-p}\,{\rm
K}^{0,0}_{p}(\tau/2)-{\rm K}^{0,0}_{p}(\tau)
\nonumber.
\label{kro}
\end{eqnarray}
Thus, Kronecker functions ${\rm K}_{2p}^{1/2,0}(\tau)$ can all be
expressed in terms of the elliptic $\theta$-functions only. For
practical calculations the following identities are also helpful
\begin{eqnarray}
 2\theta_2^2(2\tau)&=& \theta_3^2-\theta_4^2\nonumber, \\
 \theta_2^2(\tau/2)&=& 2\theta_2\theta_3  \nonumber, \\
 2\theta_3^2(2\tau)&=& \theta_3^2+\theta_4^2 \nonumber, \\
 \theta_3^2(\tau/2)&=& \theta_2^2+\theta_3^2 \nonumber, \\
 2\theta_4^2(2\tau)&=& 2\theta_3\theta_4 \nonumber, \\
 \theta_4^2(\tau/2)&=& \theta_3^2-\theta_2^2 \nonumber.
\end{eqnarray}
>From the formulas above we can easily write down  the
Kronecker functions that have appeared in our asymptotic
expansions
\begin{eqnarray}
{\rm K}_{4}^{\frac{1}{2},0}(\tau)&=&\textstyle{\frac{1}{30}
(\frac{7}{8}\,\theta_4^8-\theta_2^4\theta_3^4)}\nonumber,\\
\nonumber
\end{eqnarray}
\begin{eqnarray}
{\rm K}_{6}^{\frac{1}{2},0}(\tau)&=&-\textstyle{\frac{1}{84}
(\theta_2^4+\theta_3^4)(\frac{31}{16}\,\theta_4^8+\theta_2^4\theta_3^4)}
\nonumber.
\end{eqnarray}
Note that when $\xi\to\infty$ we have limits $\theta_2\to 0$,
$\theta_4\to 1$, $\theta_3\to 1$ . The case  $\xi \to 0$ can be obtained
by using Jacobi's imaginary transformation of the $\theta$ - functions.
In this case $\theta_2\to \frac{1}{\sqrt{\xi}}$,
$\theta_4\to 0$ and $\theta_3\to\frac{1}{\sqrt{\xi}}$  and the
Kronecker's function can again be
reduce to the Bernoulli polynomials.

\section{Asymptotic Expansion of the Digamma Function $\psi(N+\alpha)$}
\label{PsiSeries}

Let us start with well known expansion of the digamma function $\psi(N)$
\cite{GradshteinRyzhik}
\begin{eqnarray}
\psi(x)&=&\ln{x}-\frac{1}{2x}-\sum_{p=1}^{\infty}\frac{{\rm B}_{2p}}{2p}
\frac{1}{x^{2p}}
\nonumber\\
&=& \ln{x}-\sum_{p=1}^{\infty}(-1)^p \frac{{\rm B}_{p}}{p}
\frac{1}{x^{p}}.
\end{eqnarray}
Plugging in the above expansion $x=N+\alpha$ and expand the resulting factors
$\ln{(1+\alpha/N)}$, $(1+\alpha/N)^{-p}$ in powers of
$N^{-1}$ we obtain
\begin{eqnarray}
\psi(N+\alpha)&=&\ln{N}-\sum_{p=1}^{\infty}(-1)^p
\frac{\alpha^p}{p N^p}-\sum_{p=1}^{\infty}\sum_{k=0}^{\infty}
(-1)^{k+p} {\rm B}_{p} \frac{(p+k-1)!}{k!p!}\frac{\alpha^k}{N^{p+k}}
\nonumber\\
&=& \ln{N}-\sum_{p=1}^{\infty}(-1)^p
\frac{\alpha^p}{p N^p}-\sum_{l=1}^{\infty}\sum_{p=1}^{l}
(-1)^{l} {\rm B}_{p} \frac{(l-1)!}{(l-p)!p!}\frac{\alpha^{l-p}}{N^{l}}
\nonumber\\
&=&\ln{N}- \sum_{l=1}^{\infty}\sum_{p=0}^{l}
(-1)^{l} {\rm B}_{p} \frac{(l-1)!}{(l-p)!p!}\frac{\alpha^{l-p}}{N^{l}}.
\end{eqnarray}
Using the relation between Bernoulli polynomials ${\rm B}_{p}^{\alpha}$ and
Bernoulli numbers ${\rm B}_{p}$
\begin{equation}
{\rm B}_{l}^{\alpha}=\sum_{p=0}^{l}{\rm B}_{p}\frac{l!}{(l-p)!p!}
\alpha^{l-p},
\end{equation}
we finally obtain Eq. (\ref{ExpansionPsi})
\begin{equation}
\psi(N+\alpha)=\ln{N}-\sum_{p=1}^{\infty}(-1)^p
\frac{{\rm B}_{p}^\alpha}{p}\frac{1}{N^p}.
\end{equation}

\newpage
\begin{center}
{\bf FIGURE CAPTIONS}
\end{center}
\vskip 0.6cm

{\bf FIG. 1.} The Brascamp - Kunz boundary conditions for $M \times 2 N$ lattice.
 Here $M=7$ and $2N=8$.
\vskip 0.4cm

{\bf FIG. 2.} Aspect-ratio $(\xi)$ dependence of finite-size
 correction terms for the specific heat of the square lattice Ising
 model with Brascamp-Kunz boundary conditions (solid lines) and
 toroidal boundary conditions (dashed lines):
(a) $C_0$, (b) $C_1$, (c) $C_2$, and (d) $C_3$.

\newpage


\end{document}